\documentstyle[12pt]{article} 
\def\be{\begin{equation}}
\def\ee{\end{equation}} 
\def\ai{\'{\i}} 
\def\qb{\overline Q} 
\def\pb{\overline P} 
\def\pf{p_\phi}
\def\po{p_\Omega} 
\def\px{p_x} 
\def\py{p_y} 
\def\om{\Omega}

\begin{document} 
\baselineskip.33in

\centerline{\Large{\bf Path integral for minisuperspaces and its relation}}

\smallskip

\centerline{\Large{\bf with non equivalent canonical quantizations}}

\vskip1cm

\centerline{ Claudio Simeone\footnote{{\bf Electronic mail:}
simeone@iafe.uba.ar}}

\medskip \centerline{\it Departamento de F\ai sica, Facultad de Ciencias
Exactas y Naturales}

\centerline{\it Universidad de Buenos Aires, Ciudad Universitaria,
Pabell\'on I} \centerline{\it 1428, Buenos Aires, Argentina.}
\centerline{ and} \centerline{\it Instituto de Astronom\ai a y F\ai
sica del Espacio, }

\centerline{\it C. C. 67, Sucursal 28 - 1428 Buenos Aires, Argentina}

\vskip1cm

\noindent ABSTRACT

\bigskip

\noindent The relation between a recently proposed path integral for minisuperspaces  and different canonical quantizations is established. The step of the procedure where a choice between non equivalent theories is made is identified. Coordinates avoiding such a choice  are found for a class of homogeneous  cosmologies.

\vskip1cm


\newpage


\noindent
A central problem of quantum cosmology is the absence of a true time in the formalism for the gravitational field, which, in turn, includes a non physical time parameter  $\tau$ [1,2,3]. A possible solution for cosmological models can be the identification of a physical time in terms of a subset of the canonical variables describing the system under consideration [4,5,6,7]. Within this framework, in a recent book [8] we proposed a way to identify a time and obtain a consistent path integral quantization by establishing a correspondence between the action  
functional $S[q^i,p_i,N]$ of homogeneous cosmologies  and the
action ${\cal S}[Q^i,P_i,N]$ of an ordinary gauge system [9,10,11]. Canonical
gauges imposed on the gauge system define a global time for a given minisuperspace, thus yielding a quantization with a clear distinction between time and true degrees os freedom.

 A cosmological model admits our treatment if its Hamiltonian constraint $\cal H$ allows for the existence of coordinates $\tilde q^i(q^i,p_i)$ such that one of them is a global time, that is, $\{\tilde q^0,{\cal
H}\}>0$ [4], and
such that the  Hamilton--Jacobi  equation associated with $\cal H$ (or with the scaled constraint $H$) is solvable. If these conditions are fulfilled,   two succesive canonical transformations
$(\tilde q^i,\tilde p_i)\to(\qb^i,\pb_i)\to(Q^i,P_i)=(Q^\mu,Q^0,P_\mu,P_0)$ are performed, which match the  Hamiltonian
constraint $\cal H$ with the
new
momentum $P_0$; then the variables $(Q^i,P_i)$ describe an ordinary gauge system, and a canonical  gauge like  $\chi\sim Q^0-T( \tau)=0$, which does not generate Gribov copies [6,12], defines a time in terms of the variables $(\tilde q^i,\tilde p_i)$ of the model. After gauge fixation and on the constraint surface ${\cal H} =P_0=0$ the propagator takes the form [12,8] 
\be \left<\tilde q_2^i|\tilde q_1^i\right>=\int DQ^\mu
DP_\mu\exp\left(i\int_{\tau_1}^{\tau_2}\left[P_\mu
dQ^\mu-h(Q^\mu,P_\mu,\tau)d\tau\right]\right), \ee
where $h$ is the (true)
Hamiltonian for the reduced system described by the observables
$(Q^\mu (\tilde q^i,\tilde p_i,\tau),P_\mu (\tilde q^i,\tilde p_i,\tau))$. 

However, in the case of certain
cosmological models there are some subtleties which have not been
analysed in detail in our previous works, but which deserve a thorough discussion,
as they involve a definite choice between non equivalent quantizations
associated to  classically equivalent formulations; this choice is not of
the usual form associated to different operator orderings, i.e. one cannot pass from one formulation to another one by changing the operator ordering. In some
of our
works  the problem was avoided, but this was achieved
at the price of obtaining a propagator which could not be explicitly
solved [13,14].
 
The purpose of the present letter  is thus to clarify these
points and
improve the proposed procedure. First, we
shall establish which is the canonical quantization which corresponds to
our path integral formulation, and at which stage of our procedure we make
a definite choice. Within this context, we shall  analyse the positive
and negative aspects of different ways of performing  our deparametrization and quantization program.
Finally, we shall propose a coordinate change  making unnecesary the  choice between two non equivalent quantum theories in the
case of a class of minisuperspaces including, among others,  some string cosmological
models
which have recently received attention in the literature [15,16,17,13,8], as well as some anisotropic universes. Also, we shall show that our procedure is not restricted to models which admit an intrinsic time.


The discussion  can be illustrated by the first model considered in
Ref. [12], which is a flat Friedmann--Robertson--Walker universe with a
massless scalar field $\phi$ and positive cosmological constant $\Lambda$,
its scaled  Hamiltonian constraint given by
$H=-\po^2+\pf^2+\Lambda e^{6\om}= 0$, ($\om\sim\ln a$, $a$ the
scale factor). However, we shall begin by considering a 
generic constraint of the form
 \be H=-\tilde p_1^2+\tilde p_2^2+Ae^{(a\tilde q^1+b\tilde q^2)}=0 \ee with $a\neq b$. This kind of constraint
is common in dilatonic models (that is, cosmologies coming from the low
energy limit of bosonic string theory), and also includes 
isotropic
and anisotropic relativistic models, like the Kantowski--Sachs universe,
or even the Taub universe after a suitable canonical transformation [14]; the
latter is an example in which $\tilde q^0=\tilde q^0(q^i,p_i)$, so that the
time is extrinsic [18,19]. It is easy to show that a coordinate change exists such
that this constraint can be put in the form \be H=-\px^2+\py^2+\zeta
e^{2x}= 0, \ee with $sgn (\zeta)=sgn (A/(a^2-b^2))$; clearly, for $\zeta>0$ we
have a system analogous to the first studied in Ref. [12]. In the case $\zeta<0$
 the momentum  $\py$ does not vanish on the constraint surface; hence $\{p_y,H\}\neq 0$
and the coordinate $y$ is a global time . For  $\zeta>0$, instead, we have $\{p_x,H\}\neq 0$ and a global time is the coordinate $x$. 


The case involving the aforementioned subtlety corresponds to $\zeta>0$.
The straightforward application of our deparametrization and path integral
quantization procedure to the system described by the Hamiltonian (3) with
$\zeta>0$ yields, after imposing a  canonical gauge condition, a global time
$t= -x\,sgn(\px)\equiv - \eta x$ and the reduced phase space
path integral [12]
 \be \left< y_2,x_2|y_1,x_1\right>=\int DQ
DP\exp\left(i\int_{T_1} ^{T_2}\left[PdQ+\eta\sqrt{P^2+\zeta
e^{2T}}dT\right]\right), \ee
 where the endpoints are $T_1=x_1$ and $T_2=x_2$,  and the paths go from $Q_1=y_1$ to $Q_2=y_2$. 

To establish 
 which is the canonical quantization corresponding to this path
integral we can  note the following: from the reduced action yielding after gauge fixation we can read the time-depending true Hamiltonian
$h= -\eta\sqrt{p_y^2+\zeta e^{2x}}.$ Because in this case $x=-\eta  t$, then $h=-\tilde p_0=-p_x$ and we have
$$-p_x+\eta\sqrt{p_y^2+\zeta e^{2x}}=0,$$
thus obtaining two constraints, namely $K^+=0$ and $K^-=0$, each one linear in the momentum $p_x$. These two constraints  together are classically equivalent to the original Hamiltonian constraint $H=0$, which is quadratic in all the momenta; that is,  classical dynamics take place in one of two sheets in which the constraint surface splits.  
But at the quantum level this
equivalence does no more hold: a function in the kernel of the operator $\hat K^+$ or $\hat K^-$ is not annihilated by the operator  $\hat H$, but by $\hat H$ plus terms corresponding to a commutator between $\hat p_x$ and the square-root true Hamiltonian resulting from its time-depending potential (see below). It must be
emphasized that these terms cannot be eliminated by  any operator ordering; thus the choice of limiting procedure in the skeletonization of the path integral, which determines a particular ordering, is not relevant for our analysis.
Imposing the operator form of the original Hamiltonian  constraint on a wave function  yields the usual Wheeler--DeWitt equation which is
of second order in $\partial/\partial x$. Instead, splitting the
constraint into two disjoint sheets yields a canonical quantization
consisting in two equations of first order in $\partial/\partial x$: 
$$
i{\partial\over\partial x}\Psi=\pm\left(-{\partial^2\over\partial
y^2}+\zeta e^{2x}\right)^{1/2}\Psi. $$ 
In this case we have a pair of
Hilbert spaces, each one with its corresponding Schr\"{o}dinger equation.  
We can  say that the Schr\"{o}dinger quantization preserves the
topology of the constraint surface, that is, the splitting of the
classical solutions into two disjoint subsets has its quantum version
in the splitting of the theory into two Hilbert spaces.
  
Therefore, our path integral quantization is not in correspondence with
the solutions of a Wheeler--DeWitt equation, but with those of a pair of
Schr\"{o}dinger equations, one for each sheet of the constraint surface (we want to insist on the point that operator ordering plays a secondary role within this context).
 Though the Wheeler--DeWitt equation is the most common choice for the
canonical  quantization of minisuperspaces,  a time-depending potential in
the Hamiltonian constraint  makes difficult the
interpretation of the resulting wave function in tems of a conserved
positive-definite inner product  (this  cannot be  avoided 
 by defining the time as $\pm y$, because when $\zeta >0$ we do not have
$\{y,H\}\neq 0$ everywhere; such a  wrong choice would lead to a reduced Hamiltonian which is not self-adjoint).  The Schr\"{o}dinger quantization, instead,
allows to define a conserved inner
 product for each subset of solutions associated to each sheet of the
constraint; this  inner product is defined by fixing the time
in the
integration:
 $$ (\Psi_1|\Psi_2)=\int dq\,\delta
(t-t_0)\Delta\Psi^*_1\Psi_2, $$
where $\Delta$ is a determinant making
the integral independent of the time choice (if one of the coordinates $q$
explicitly appearing in the wave function is itself the time, then $\Delta=1$). Hence  our
path integral formulation is associated to  the canonical quantization
which allows for a clear probability interpretation, and which, in a sense, reproduces the classical geometry of the   constraint surface. 
Our  procedure could  
 be understood as introducing the appropriate quantum 
corrections to the Hamiltonian constraint $H$ [20], which are given by  the   commutator mentioned above, and whose general form is  $\left[\sqrt{\sum(\hat{\tilde p_r})^2 +V(\hat{\tilde q^i})},\hat{\tilde p}_0\right]$ (where $r\neq 0$, and $V$  stands for the potential in the scaled Hamiltonian constraint $H$). 

The choice between both formalisms is made when we solve the
Hamilton--Jacobi equation $H\left(\tilde q^i,\partial W/\partial \tilde q^i\right)=E$: because of the quadratic form of the constraint, this is  a first order non linear equation, but  to explicitly obtain its solution $W$ one integrates
two equations which are linear in the derivative respect to the coordinate
identified as $\pm$ the time,
$$
{\partial W\over\partial \tilde q^0}=\pm\left(\sum_r\left({\partial W\over\partial\tilde q^r}\right)^2 +V(\tilde q^i)-E\right)^{1/2}.$$
  At the level of the path integral this is reflected in the fact that the reduced Hamiltonian $h$ in (1) is equal to $\partial f/\partial \tau$, where $f$  is chosen  to ensure that the path integral in the new variables effectively corresponds to the  transition amplitude $\langle \tilde q^i_2\vert\tilde q^i_1 \rangle$. This is achieved if  the end point terms [11,12] 
$$B= \left[ \overline Q^i\overline P_i -W(\tilde q^i,\overline P_i)+Q^\mu P_\mu-f(\overline Q^\mu ,P_\mu ,\tau)\right]_{\tau_1}^{\tau_2}$$
associated to the two succesive canonical transformations vanish on the constraint surface $H=0$ and in a gauge associated to a global time $t(\tilde q^i)$;  this yields two disjoint theories, one for each   reduced Hamiltonian determined by  each sign of $W$ .

In some of our works [13,14] we proposed a change to null coordinates leading
to the following form of the constraint: 
\be  p_x p_y+C=0. \ee 
The
obtention of a constant `potential' then seems  to solve the problem
associated to the existence of non equivalent forms of writing the
constraint. The application of our 
method starting from this constraint leads to an action including a true
Hamiltonian which is time-independent. However, the resulting propagator
for the physical degree of freedom 
\be \left< y_2,x_2|y_1,x_1\right>=\int
DQ DP\exp\left(i\int_{T_1} ^{T_2}\left[PdQ-{\eta\over P}dT\right]\right)
\ee 
(where the paths go from $Q_1=y_1$ to $Q_2=y_2$ and the endpoints are
$T_1=x_1,\ T_2=x_2$) has the unsatisfactory
feature that the functional integration over $P$  cannot be effectively performed, even to find its
infinitesimal form, because of the dependence $P^{-1}$ of the reduced
Hamiltonian.


For the class of models studied here we can  propose a
coordinate choice with the following  desirable properties:
1) The necessity of deciding between inequivalent
quantum theories yielding from different forms of writing the classical constraint is avoided. 
2) The functional
integration  can be explicitly performed.
Consider the constraint (2) and define 
\begin{eqnarray} u& =&\alpha
\exp\left({a\tilde q^1+b\tilde q^2\over 2}\right)\cosh \left(b\tilde q^1+a\tilde q^2\over 2\right)\nonumber\\ v&
=&\alpha \exp\left({a\tilde q^1+b\tilde q^2\over 2}\right)\sinh \left(b\tilde q^1+a\tilde q^2\over 2\right),
\end{eqnarray} 
with $\alpha=\sqrt{|A|}$. Because $a\neq b$ and $u\neq v$ these coordinates allow to write the constraint in the equivalent (scaled) form 
\be H=
-p_u^2+p_v^2+\eta m^2=0, \ee with 
$\eta=sgn (A)$ and
$m^2=4/|a^2-b^2|$. It is clear that  commutators cannot appear now, or, in other words, the Wheeler--DeWitt equation is equivalent to two Schr\"{o}dinger equations. The time will be  $u$ or $v$  depending on $\eta$; formally, this yields from a canonical gauge of the form $\chi\equiv\sqrt{P^2+m^2}Q^0-T(\tau)=0$.  The application of our  path integral procedure  to this system is straightforward; on the constraint surface and after gauge fixation we obtain:
\be
\langle u_2,v_2\vert u_1,v_1\rangle =\int DQ\,DP\exp \left(i\int_{T_1}^{T_2} \left[PdQ+\eta\sqrt{P^2+m^2}dT\right]\right),
\ee
where the endpoints are $T_1=u_1(v_1)$ and $T_2=u_2(v_2)$ and the paths go from $Q_1=v_1(u_1)$ to $Q_2=v_2(u_2)$ for $\eta= 1(-1)$.  By skeletonizing the paths we obtain $N-1$ $\delta$-functionals of the form $\delta(P_m-P_{m-1}),$ and  hence the functional integration reduces to an ordinary one:  
\be 
\langle u_2,v_2\vert u_1,v_1 \rangle =\int dP\exp\left( i\left[  P(Q_2-Q_1)+\eta \sqrt{P^2+m^2}(T_2-T_1)\right]\right). 
\ee  
The double sign given by $\eta$  corresponds to both possible sheets of the constraint surface where the evolution can take place.

 Let us illustrate this coordinate choice with some simple dilatonic  cosmologies (see [8] and references therein); consider the scaled  constraint
$$H=-\po^2+\pf^2+2ce^{6\om+\phi}=0$$
which corresponds to a flat  model with dilaton field $\phi$. In the case $c<0$ this constraint admits a change to the coordinates $x$ and $y$ yielding an expression like (3), with $t=\pm y$, but for $c>0$ we have $t=\pm x$ and the problem of the time-depending potential appears.  The change to $(u,v)$, instead, solves this: for $c<0$ we have $t=\pm v$, while for $c>0$ we obtain $t=\pm u$. Note that in the case $c<0$ (for which the dilaton $\phi$ is itself a globally good time as $p_\phi\neq 0$),  we obtain $-\infty<t<\infty$ on boths sheets of the constraint determined by the sign of $p_v$; in the case $c>0$ (which admits $\om$ as a global time), instead, we have that $t$ goes from $-\infty $ to $0$ on the sheet $p_u>0$, and from $0$ to $\infty$ on the sheet $p_u<0$, with $t\to 0$ corresponding to the singularity $\om\to -\infty$. If we add a term representing the inclusion of  a non vanishing antisymmetric field $B_{\mu\nu}$  coming from the $NS$-$NS$ sector of effective string theory, we have the constraint
$$H=
-\po^2+\pf^2+2ce^{6\om+\phi}+\lambda^2e^{-2\phi}= 0
$$
which in principle does not admit the proposed coordinate change. Moreover, in the case $c<0$ the model does not admit an intrinsic time. However, we should recall that, because  these models come from the  low energy string theory, which makes sense in the limit $\phi\to -\infty$, then     the $e^\phi\equiv V(\phi)$ factor in the first term of the potential verifies $V(\phi)=V'(\phi)\ll 1$ (this is clearly not the case with the term $e^{-2\phi}$), and  we can replace $ce^\phi\to \overline c$. As a previous step we can then perform the canonical transformation introduced for the Taub universe in Refs. [7,14] to obtain  a constraint with only one term in the potential: for both signs of $\overline c$ we can define the generator $f_1=\pm |\lambda|e^{-\phi}\sinh s$ of a canonical transformation leading to 
$$H=-p_\om^2+p_s^2+2\overline c e^{6\om}=0$$
and we can apply our procedure starting from this constraint. As before, for $\overline c<0$ we obtain $t=\pm v$, while for $\overline c>0$   we obtain $t=\pm u$. Note that now there is an important new feature: because both $u$ and $v$ depend on the `intermediate' coordinate $s$ which involves in its definition the original momenta, the time is extrinsic, that is, $t=t(q^i,p_i)$ (in the case $\overline c<0$ an intrinsic time does not exist). However, in the case of interest for us, which is $\overline c>0$, we have the same behaviour of $t$ with $\om$ that we had in the absence of the antisymmetric field; $t$ goes from   $-\infty$ to $0$ on the sheet $p_u>0$ of the constraint surface and from $0$ to $\infty$ on the other sheet, while $t\to 0$ for the singularity $\om\to -\infty$. 

Though the problem of time in quantum cosmology is far from having been solved, we believe that at the level of the minisuperspace approximation much can be done towards a consistent formulation, both within the path integral formalism and in the usual canonical formalism. Here, within the context of the problem of non equivalent quantizations yieldind from the same classical theory, we have discussed essentially three points: the correspondence between our path integral approach and a Schr\"{o}dinger equation, the step of our deparametrization procedure where the choice takes place, and finally the possibility of avoiding the necessity of deciding between non equivalent quantizations by means of a suitable  coordinate choice. Of course, the latter works for a limited  class of homogeneous models, but, as we could see, it includes both string and relativistic, isotropic and some anisotropic,  cosmologies; also, we  have seen that we can even deal with models which do not admit an intrinsic time. 

The possibility of working with an extrinsic time suggests  an alternative formulation in the line of Ref. [21]. There, we avoided the intermediate coordinates $\tilde q^i$ by a straightforward procedure which led to  a quantization in which the states were not characterized by the coordinates, but by the momenta; thus we obtained a  path integral quantization  with a good global phase time for the closed ($k=1$) de Sitter universe. Working with an extrinsic time allowed to avoid the splitting of the formulation into two disjoint theories, even when an intrinsic time existed (see the case of the open ($k=-1$) de Sitter universe in the same paper); an attemp to  generalize this analysis  could be an interesting line of work to be followed.

\vskip1cm

\noindent {\Large{\bf Acknowledgements}}

\bigskip

\noindent This work was supported by CONICET and UBA (Argentina). 

\newpage

\noindent {\Large{\bf References}}

\bigskip

\noindent 1.  J. J. Halliwell, in {\it Introductory Lectures on Quantum Cosmology}, Proceedings of the Jerusalem Winter School on Quantum Cosmology and Baby Universes, edited by T. Piran, World Scientific, Singapore (1990). 

\noindent 2.  A. O. Barvinsky, Phys. Rep. {\bf 230}, 237 (1993).

\noindent 3. J. J. Halliwell, gr-qc/0208018.

\noindent 4. P. H\'aj\'{\i}cek, Phys. Rev.  {\bf D34}, 1040 (1986).

\noindent 5. S.  C. Beluardi and  R.  Ferraro, Phys. Rev.  {\bf D52}, 1963 (1995).

\noindent 6. C. Simeone, J. Math. Phys. {\bf 40}, 4527 (1999).
 
\noindent 7. G. Catren  and R.  Ferraro, Phys. Rev. {\bf D63}, 023502 (2001).

\noindent 8. C. Simeone, {\it Deparametrization and Path Integral Quantization of Cosmological Models}, World Scientific Lecture Notes in Physics 69, World Scientific, Singapore (2002).

\noindent 9. M. Henneaux, C.  Teitelboim  and  J. D. Vergara, Nucl. Phys.  {\bf B387}, 391 (1992).

\noindent 10.  M. Henneaux  and C. Teitelboim, {\it Quantization of Gauge Systems,} Princeton University Press, New Jersey (1992).

\noindent 11. H. De Cicco  and   C. Simeone,  Int. J. Mod. Phys. {\bf A14}, 5105 (1999).

\noindent 12. C. Simeone, J. Math. Phys. {\bf 39}, 3131 (1998).

\noindent 13. G. Giribet  and  C. Simeone, Mod. Phys. Lett. {\bf A16}, 19  (2001).

\noindent 14. G. Giribet  and  C. Simeone, Int. J. Mod. Phys. {\bf A17}, 2885 (2002).

\noindent 15. M. Cavagli\`a   and C.  Ungarelli, Class. Quant. Grav. {\bf 16}, 1401 (1999). 
 
\noindent 16. M. Gasperini,  Int. J. Mod. Phys. {\bf A13}, 4779 (1998).

\noindent 17. M. Gasperini and G. Veneziano, hep-th/0207130.

\noindent 18. J. W.  York, Phys. Rev. Lett. {\bf 28}, 1082 (1972).

\noindent 19.  K. V. Kucha\v r, J. D. Romano  and M.  Varadajan,    Phys. Rev.  {\bf D55}, 795 (1997).

\noindent 20. S. Carlip, Class. Quant. Grav. {\bf 11}, 31 (1994).

\noindent 21. G. Giribet  and  C. Simeone, Phys. Lett. {\bf A287}, 344   (2001).

\end{document}